\documentclass[showpacs,amsmath,amssymb,twocolumn,aps,
        superscriptaddress]{revtex4-1}

 \usepackage{graphicx}% Include figu1re files
\usepackage{epsfig}             
 \usepackage{pstricks}

 \usepackage{pst-plot}
  \usepackage{psfrag}
 \usepackage{dcolumn}% Align table columns on decimal point
 \usepackage{bm}% bold mat

\usepackage{color}

\def\mn#1{}

\begin{document} 
\title{Phonon induced Superconductivity of High Temperatures in Electrical 
Graphene Superlattices}
\textsc{}
\author{J\"urgen Dietel}
% \email{dietel@physik.fu-berlin.de}
\affiliation{Institut f\"ur Theoretische Physik,
Freie Universit\"at Berlin, Arnimallee 14, D-14195 Berlin, Germany}

\author{Victor Bezerra}
% \email{dietel@physik.fu-berlin.de}
\affiliation{Institut f\"ur Theoretische Physik,
Freie Universit\"at Berlin, Arnimallee 14, D-14195 Berlin, Germany}

\author{Hagen Kleinert}
% \email{kleinert@physik.fu-berlin.de}
\affiliation{Institut f\"ur Theoretische Physik,
Freie Universit\"at Berlin, Arnimallee 14, D-14195 Berlin, Germany}
\affiliation{ICRANeT, Piazzale della Repubblica 1, 10 -65122, Pescara, Italy}
\date{Received \today}

\begin{abstract}
We discuss the BCS theory for electrons in graphene with a superimposed 
electrical unidirectional superlattice potential (SL). New Dirac points 
emerge together with van Hove singularities (VHS) linking them.   
We obtain a superconducting transition temperature $ T_c $ 
for chemical potentials close to the VHS assuming that 
acoustic phonon coupling should be 
the dominant mechanism. Pairing of two onsite electrons with one electron  
close to the $ {\bf K} $ and the other close to the  
$ -{\bf K} $ point is the most stable pair formation.   
The resulting order parameter 
is almost constant over the entire SL.            
\end{abstract}


\maketitle

\section{Introduction} 
The emergence of new interesting physics by the application 
of electrical and magnetical fields on graphene is one of the  
properties of this material.  
It was shown for example recently that new Dirac points 
can be opened in the energy spectrum by imposing an electrical   
SL on the graphene layer \cite{Talyanskii1, Brey1, Park2}. 
Most important in neutral graphene is 
that these new Dirac points are opened up in the lowest energy 
band. Other 
Dirac points emerge as linking points of two minibands 
\cite{Park1,Park5} at higher energies. 
New Dirac points were in fact found experimentally 
for graphene with Moir\'{e} SLs on underlying substrates 
 \cite{Pletikosic1,Yankowitz1} and in unidirectional 
corrugated graphene monolayers \cite{Yan1}. 
Such points lead to unusual conductivity properties in SL systems 
\cite{Brey1, Park2, Dietel2, Dietel3, Burset1,Barbier2,Sun1}. 
Together with the new Dirac points also VHSs emerge in the density of states 
shown up as saddle points in the energy spectrum.  
The new Dirac points are linked by the saddle points.  

Since the discovery of graphene there were attempts to find superconductivity 
in these materials. This is mainly motivated by the fact that 
superconductivity shows up experimentally in other carbon based 
materials    
with rather high critical temperatures $ T_c $ for 
conventional superconductors as for example graphite intercalated 
($ T_c \lesssim 12 K $) \cite{Weller1, Calandra1} 
and fullerite compounds ($ T_c \lesssim 33K $) 
\cite{Gunnarson1}. Both forms of carbon based superconductors are mainly well 
described by the conventional phonon mediated BCS theory. 
The higher temperatures in the fullerite superconductors can be attributed 
to the  high frequency of the 
innermolecular phonon modes being responsible for pairing 
in fullerites \cite{Varma1}. These phonon modes have around one order of 
magnitude higher 
frequencies than  phonons mediating BCS superconductivity in metals 
\cite{Eisenstein1}. Similar high phonon frequencies are also found 
in the graphene phonon spectrum. Furthermore, theoretically  it was shown 
that also graphane \cite{Savini1}, multilayer \cite{Kopnin1} 
and strained graphene \cite{Uchoa2}
could lead to BCS instabilities with high temperatures. 
In Refs.~\onlinecite{Uchoa1, Kopnin2} it was shown theoretically
that for pristine graphene at half filling  
a critical interaction value exists  
above which BCS pairing is possible. 
This is mainly due to a vanishing density of states at half filling. 
In both papers restrictions on the electronic pairing are made 
where either a coupling with total zero momentum \cite{Uchoa1} is considered  
or, more restrictively with an onsite s-wave pairing 
of one electron close to the   
$ {\bf K} $ with another electron close to the  $ {\bf -K} $ valley 
\cite{Kopnin2, Uchoa2}. For small but non-zero chemical potentials 
gained by electrostatic doping, $ T_c $ is still small. 
Except of the small density of states at these fillings 
one has to take also into account here the smallness of the optical 
electron-phonon coupling constant which is relevant in this regime 
\cite{Einenkel1}. 
The corresponding deformation potential for the coupling of electrons with  
longitudinal acoustic $ {\bf \Gamma} $ phonons is much higher 
than of the other acoustic and optical phonon modes \cite{Suzuura1, Ishikawa1, 
Pisanec1, Dietel4}.  
This coupling mechanism should become relevant for 
larger chemical potentials when the corresponding 
Bloch Gr\"uneisen temperature $ \Theta_{\rm BG} 
=2  \hbar k_F v_s $ in which   $ k_F $ is the Fermi-momentum and $ v_L $ the 
phonon velocity, is in the regime of the Debye temperature \cite{Efetov1}. 
Here we use the fact that   
$ T_c $ scales exponentially with the inverse square of the 
electron-phonon coupling, but only factorially with the energy  
cut-off $ \sim \Theta_{\rm BG}/2 $ for acoustic phonon coupling,  or 
the main  optical phonon frequency for optical phonon pairing. 
Note that in graphene the Debye frequency of the 
longitudinal acoustic $ {\bf \Gamma} $ 
phonons is of similar magnitude  as of the main optical phonons.     
With the application of a SL, the electron bands are 
effectively folded bringing the effective Gr\"uneisen temperature 
also for low electrostatic doping potentials 
in the regime where the deformation 
potential coupling becomes relevant. 
This is one motivation to consider superconductivity in graphene 
superimposed by an SL.  An additional motivation is the existence of low-lying  
VHSs in SL graphene which promises superconductivity with 
high $ T_c $-values for chemical potentials close to the VHSs.   
 
There are other possible sources of superconductivity than 
only phonon mediated superconductivity. One finds in the literature  
for example the Coulomb interaction as a possible  
source of pairing via the Kohn-Luttinger mechanism in graphene 
\cite{Gonzales1, Nandkishore1, Kiesel1}.
This effect becomes most pronounced for energy bands when a VHS 
is existent.  In pristine 
graphene one finds  three in-equivalent saddle points 
producing a VHS  at large energies linking the 
$ {\bf K} $ and $- {\bf K} $ Dirac points. Such high chemical potentials 
can yet only be reached by chemical doping \cite{McChesney1}. 
It was shown in Ref.~\onlinecite{Gonzales1} that a possible $ d+id $ 
wave instability with high $ T_c $ can only be guaranteed 
when the saddle points producing  
the VHS are linked approximatively by nesting vectors. 
Such nesting vectors are not found  for the VHSs in SL systems. 
Note that phonon-coupled BCS theory is not yet discussed for 
high chemical doped graphene in the literature. 
One reason is that phonon modes are sensitive on the special 
chemical doping which makes it rather complicated to 
carry out such calculations \cite{Calandra1}.   
 
In the following, we will discuss the simplest case of BCS-type 
superconductivity in SL superimposed graphene 
mediated by acoustic phonons. We concentrate us hereby to the most interesting 
region of  chemical potentials close to  
VHSs since this promises the highest $ T_c $-values. 
Since we shall use analytically the role of the different 
possible superconducting order parameters in the SL system, 
our investigation can in principle be used  
when other superconducting coupling mechanisms become relevant. 

The paper is structured as follows. 
In Sect.~II we give first an  introduction to the Bogoliubov-de Gennes 
(BdG) equation 
for superconductivity in graphene superimposed
with an SL and discuss the transfer matrix formalism for solving this equation. 
Sect.~III discusses the one-particle spectrum, and 
Sect.~IV the phase diagram as a function of temperature.

\section{Electrical Superlattice}

In the following we neglect corrections to BCS superconductivity 
expressions due to the repelling Coulomb interaction. 
Here we take into account that the unscreened  
interaction potential 
of electrons due to Coulomb interaction is of similar value  
as the attractive interaction potential from the Fr\"ohlich 
Hamiltonian (c.f. Eq.~(\ref{15})) for momentum transfer 
$ k_{\rm DB} \approx 2.2/a $ calculated 
by using longitudinal acoustic electron-phonon coupling, 
where $ k_{\rm DB} $ is the Debye momentum and 
$ a \approx 1.4 $\AA~ the interlattice distance.
Due to the large momentum transfer, 
we can neglected in our calculation screening effects 
due to a possible substrate and further the inner graphene screening.   
For electron band widths much larger than the energy cut-off 
due to the electron-phonon interaction, retardation 
effects becomes important and an electron scatters  
with the phonon trace of another electron being not close in space
at the same time \cite{Gunnarson1,Alexandrov1}.  
This leads to a suppression of the effective Couloumb 
interaction potential known as the 
so called Coulomb pseudopotential. This potential is strongly  
suppressed for superconductors where 
the density of states is large at the Fermi-surface \cite{Morel1,Sigrist1}.  
This is the case in the regime we are interested in when the chemical 
potential of the SL system lies close to a VHS. 
For small momentum transfer we can neglect the Coulomb   
interaction due to the large screening in the vicinity of the 
VHSs.   
 
We discuss  here the  most simple representation of a SL  
being a symmetric  two-step Kronig-Penney potential with 
a superlattice potential $ V(x)=V \chi(x) $ where 
 $ \chi(x)={\rm sg}[\sin(2 \pi x/d)]  $. 
The function $ {\rm sg}[x] $  is the 
sign of $ x $, and $ d $ is 
the wavelength of the SL.  
In the continuum approximation,  
the graphene Hamiltonian under consideration near the  
$ \pm {\bf K} $ Dirac point is    
given for $ d \gg a $ by \cite{Castro1}
\begin{equation} 
H_{\pm}=\hbar 
v_F (\pm \sigma_1 \partial_{x}/i + \sigma_2 \partial_y/i) 
+ V(x)    \,.   \label{5} 
\end{equation}  
 Here $ \sigma_{1,2} $ 
are the Pauli matrices, while $ v_F $ is the  
velocity of the electrons in graphene. In the following, we assume as in 
conventional superconductors  spin singlet pairing, being most reasonable 
for phonon pairing.
The formalism is then simplified considerably by 
taking into account the eigenvalue problem in the Nambu space with 
the eight component field  
$ {\bf \Psi }(x)=(\phi^K_{A,\uparrow} \! , \! 
\phi^K_{B,\uparrow}\! ,\! -i \phi^{-K}_{B,\uparrow} \! , \! 
i \phi^{-K}_{A,\uparrow} \! , \! (\phi^{-K}_{A,\downarrow})^* \! , \!  
(\phi^{-K}_{B,\downarrow})^* \! , \!  i (\phi^{K}_{B,\downarrow})^* \! , \! 
-i (\phi^{K}_{A,\downarrow})^*) $. 
The BdG-Hamiltonian is given by 
\begin{equation} 
{\bf H}_{\rm BdG}= \left( \begin{array}{c c}  \sigma_0 \otimes 
(H_+ -\mu)  &  \Delta \\
                       \Delta^+   & -\sigma_0 \otimes (H_+-\mu)    
             \end{array} \right) \label{10}
\end{equation}   
where  $ \sigma_0 $ is the two-dimensional unit matrix.      
The condensate matrix $ \Delta_{ij} $ is given by   $ \Delta_{ij}({\bf r})= 
(g/S^2) \sum_{{\bf k},{\bf q}} 
 \langle \Psi_i ({\bf k}+{\bf q}) \Psi^+_{4+j}({\bf k})\rangle  
\theta({\bf k}+{\bf q}) \theta({\bf k}) 
 e^{i{\bf q}{\bf r}} $
where $ S $ is the area of the system. 
The function $ \theta({\bf k}) $ is an energy  cut-off given 
by $ \theta({\bf k})\equiv 
\Theta[\omega_{\rm DB}^* -|\epsilon^0({\bf k}|)] $ for 
some canonical momentum $ {\bf k} $ where $ \Theta $ 
is the Heaviside function. Here $ \epsilon^0({\bf k}) $ is the energy of 
the lowest band of (\ref{10}) 
for $ \Delta=0 $ (c.f. Eq.~(\ref{80}) below). We point out 
that $ {\bf k} $ and $ {\bf k}+{\bf q} $ are canonical momenta and 
not the Bloch momenta of the eigenfunctions. A sufficient condition 
that the BdG equation (\ref{10}) is then a  
mean-field BCS decoupling equation   
for the exact superconducting problem by using 
the Fr\"ohlich interaction approximation  
requires that the eigenfunctions of (\ref{10}) for $ \Delta_{ij}=0 $ are   
localized  on a circle in canonical momentum space 
for electrons with energies close to the chemical potential. 
That (\ref{10}) together 
with $ \Delta_{ij} $ in the canonical momentum basis is well defined 
requires further that the energy band $ \epsilon^0({\bf k}) $ 
is a unique function of the canonical momenta.  
Both assumptions will be shown below  
where we also determine the energy cut-off $ \omega_{\rm DB}^* $. 

Let $ g $ denote  
the phonon induced coupling constant of the Fr\"ohlich 
Hamiltonian for graphene.  
Due to the inhomogeneity  of the SL in 
space it is not appropriate to consider a constant pairing function.
Instead we shall assume an order parameter which is step-like of the form 
$ \Delta_{ij}({\bf r})= \Delta^c_{ij}+ 
\Delta^s_{ij} \chi(x) $,  where $ \Delta^c_{ij}$ and $ \Delta^s_{ij} $ are 
constant.
% This is justified by the fact that we consider 
% here the case that the chemical potential is in the vicinity of a 
% VHS where the relevant momentum region in the contribution to the gap function% is restricted considerably. 

The acoustic electron-phonon energy due to deformation potential 
coupling  is given by 
$ H_{\rm ep} = g_{\rm ep} \int d{\bf r}  (u_{xx}+u_{yy}) 
|{\bf \Psi}({\bf r})|^2  $ where 
$ g_{\rm ep} $ is the deformation potential and $ u_{ij} $ is the strain tensor 
of the graphene lattice.  
The effective Fr\"ohlich interaction coupling constant $ g $ is then 
given by $ g = g_{\rm ep}^2 2 /v_L^2 \rho_C $ where 
$ v_L \approx 21.1 \cdot 10^3 m/s$ 
is the longitudinal acoustic phonon velocity, 
and $ \rho_C \approx 761 \cdot 10^{-9} kg/m^2 $ the density of carbon atoms. 
This leads to $ g \approx  6 \cdot10^{-19} m^2 \rm eV $.
Here we work 
with a  deformation potential of $ g_{\rm ep} \approx 25 \rm {\rm eV} $.
The corresponding Fr\"ohlich 
coupling constant for out-of-plane acoustic phonons 
is by a 
factor $ \omega_{\rm DB}^2/\kappa_0^2 \ll 1 $ smaller, where $ \kappa_0 $ 
is the bending constant \cite{Dietel1} and $ \omega_{\rm DB} $ is the 
Debye frequency for longitudinal acoustic phonons.  
The Fr\"ohlich interaction Hamiltonian is then 
\begin{align} 
& H_{\rm Fr}\! = \! -\frac{g}{S^3} \! \! \sum_{i,j \le 4}  
\sum_{{\bf k}, {\bf k}',{\bf q}} \! \! \! \! 
\Psi_{4+j}(\!{\bf k}\!) \Psi^+_i (\! {\bf k}+{\bf q}\! ) 
\Psi_i (\!{\bf k}'+{\bf q}\!) \Psi^+_{4+j}(\!{\bf k}'\!) \nonumber \\
& \qquad \qquad \times \theta({\bf k}) \theta({\bf k}+{\bf q}) 
\theta({\bf k}'+{\bf q}) \theta({\bf k}')\,.  \label{15}  
\end{align}   
           
We obtain from (\ref{5}) and (\ref{15}) by using a mean-field decoupling 
the BdG Hamiltonian (\ref{10}) where the BdG-matrix has then in 
general ten  unknown complex 
parameters $ \Delta_{ij}  $. Here we assume  
that $ \Delta_{ij} = (-1)^{i+j+1}
\Delta_{4-i,4-j} $ for $ i,j \in \{1,2\} $,  and $ \Delta_{31}=-\Delta_{42} $, 
$ \Delta_{13}=-\Delta_{24} $
when we take into account spin singlet pairing 
in the original graphene fields. 
One can simplify this 
matrix further under the assumption that 
the condensate does not break 
the time-inversion symmetry as well as the mirror symmetry with respect 
to the $x$ and $y $-axis, 
where we choose that the mirror 
operation with respect to the $x$-axis should lead to an 
interchanging of $A$, $B$ atoms if $ A$ and $B$ denotes the 
inequivalent carbon atoms in the fundamental cell. 
These assumptions will be justified further below. 
The time inversion transformation on a graphene spinor is given by 
$ {\cal T} \phi^{K}_{A,\uparrow}= (\phi^{-K}_{A,\downarrow})^* $, defined 
modulo the interchange $ A \leftrightarrow B $, 
$ \uparrow \leftrightarrow \downarrow $
and $ K \leftrightarrow -K $. Similar is also assumed 
for the $x$-axis mirror transformation 
$ {\cal S}_x \phi^{K}_{A,\uparrow}(y)= \sigma_1 \phi^{K}_{B,\uparrow}(-y)$ and 
the $y$-axis mirror transformation 
$ {\cal S}_y \phi^{K}_{A,\uparrow}(x)= \phi^{-K}_{A,\uparrow}(d/2-x)$.
By taking into account the invariance of the condensate under these operations, 
we obtain $ \Delta= \Delta_1+\Delta_2 + \Delta_3 + \Delta_4$ with 
\begin{align} 
& \Delta_1= d_1 \sigma_3 \otimes \sigma_0 \,, \,   
\Delta_2= d_2 \sigma_0  \otimes \sigma_1 \,, \,    \nonumber \\
&  \Delta_3  =  d_{3} \sigma_2  \otimes  \sigma_2 \, 
, \,   \Delta_4= d_4 \sigma_1  \otimes  \sigma_3 \,,   \label{20}
\end{align}
where $ d_i \in \mathbb{R}$. We now separate $ d_{i} $ according to 
$  d_{i}=d_i^c+ d_i^s \chi(x) $ where $ d_i^c$, $d_i^s $ are constants.   
In the following, we solve the eigenvalue 
equation  $ {\bf H}_{\rm BdG } {\bf u}({\bf r}')=\epsilon \,  
{\bf u}({\bf r}') $ by 
using the transfer matrix method \cite{Arovas1, Dietel2}.
With the help of $  {\bf u}(x,y) \! = \! e^{i k_y y} {\bf u}(x) $, 
the eigenfunctions of the lowest band   
are given by   
$  {\bf u}(x)= {\bf \Lambda}(x) {\bf u}(0) $. 
With this definition   
we obtain from the Schr\"odinger equation with the Hamiltonian (\ref{10}) 
the following equation 
for the transfer matrix $ {\bf \Lambda} $ 
\begin{align} 
& \frac{1}{i} \partial_x {\bf \Lambda}(x)\! = \! - 
\! \sigma_3 \! \otimes \! \sigma_0 \! \otimes \! 
\sigma_3 
\big(k_y \sigma_3 \! \otimes \! \sigma_0 \! \otimes \! \sigma_2 \! + \! V(x) 
\sigma_3\!  \otimes \! \sigma_0 \! \otimes \! \sigma_0   \nonumber \\
& 
\qquad \qquad \; \; \,   - \epsilon
\sigma_0 \otimes \sigma_0 \otimes \sigma_0 + \sigma_1 \otimes \Delta \big){\bf \Lambda}(x)\,. \label{30}  
\end{align}        

This equation is solved 
perturbatively with respect to the small  condensate matrix 
$ \Delta $, where the corresponding terms are denoted by 
$ {\bf \Lambda}= {\bf \Lambda}^0+{\bf \Lambda}^1+{\bf \Lambda}^2+\ldots $. 
We obtain from (\ref{30}) that for $ \Delta=0 $, 
$ {\bf \Lambda}={\bf \Lambda}^0 $ 
is diagonal within the valley and electron-hole sectors.
We denote the valley electron-hole submatrizes by    
$ \Lambda^0_\pm= \sum_i \sigma_i {\rm Tr}
[(\sigma_0 \pm \sigma_3) \otimes 
\sigma_0 \otimes (\sigma_i)^* \cdot  {\bf \Lambda}^0]/4$. This leads to

\begin{equation} 
\Lambda_{\pm}^0(x)\!  = \! \lambda_0(x) 
\Theta \! \left(\! \frac{d}{2}-x \! \right) \! + \!  
\lambda_{d/2}(x) \lambda_{0}\! \left(\!\frac{d}{2}\!\right) \! 
\Theta\!\left(\! x- \frac{d}{2}\! \right) \,,  \label{40} 
\end{equation}  
where   
\begin{align} 
&  \lambda_{x_0}(x) \!  = \!   \cos\! \left[\!   
\frac{\alpha_{E_\pm}(x) 2 (x-x_0)}{d} \! \right]\! \!  \sigma_0 \! + \!  
\frac{\sin \! \left[ \! 
\frac{\alpha_{E_\pm}(x) 2 
(x-x_0)}{d}\! \right] \!}{\alpha_{E_\pm}(x)}  M_\pm  \label{50} 
 \end{align} 
with  
\begin{equation} 
M_\pm= k_y \sigma_{3} +[E_\pm -V(x)] \sigma_2/\hbar v_F\,. 
\label{60}     
\end{equation}
Here $ E_\pm= \pm \epsilon + \mu $ and 
\begin{equation} 
 \alpha_{E_\pm}(x)  = \{[(E_\pm-V(x))/\hbar v_F]^2 -
      k_y^2\}^{1/2} d/2 \,.   \label{55} 
\end{equation} 

We can now calculate the energy spectrum 
for $ \Delta =0 $ by using the Bloch condition 
\textsc{}\begin{equation} 
 \Lambda_\pm^{0}(d) u_\pm^0(0) = e^{i k_x d} u_\pm^0(0)  \label{70} 
\end{equation}    
which is effectively an eigenvalue equation for  
$ \Lambda^{0}_\pm(d) $ where the Bloch condition demands that the 
eigenvalue is a phase. By using  
the mirror symmetry of the SL with respect to the  axis $ x=d/4 $, 
we obtain that eigenvalues of the transfermatrix $ \Lambda^0_\pm(d) $ to 
the Hamiltonian must come in pairs $ e^{ik_xd} $ and $ e^{-ik_xd}$,  where 
$ k_x $ and $ -k_x $ are complex numbers in general. In the case of the 
Bloch eigenvalue equation (\ref{70}) 
this leads to $ {\rm Tr}[\Lambda_\pm^{0}(d)]=2\cos(k_x d) $, where 
\begin{align} 
& {\rm Tr}[\Lambda_\pm^{0}(d)]= 2 \cos[\alpha_{E_\pm}(d/4)] 
\cos[\alpha_{E_\pm}(3d/4)]  \nonumber \\
& -
2 \frac{\sin[\alpha_{E_\pm}(d/4)] 
\cos[\alpha_{E_\pm}(3d/4)]}{\alpha_{E_\pm}(d/4)\alpha_{E_\pm}(3 d/4)}
[\tilde{k}_y^2-(\tilde{E}_\pm^2-\tilde{V}^2)] .  \label{75}  
\end{align}

For the energy dispersion  in the 
lowest band,  we obtain for large SL potentials $ \alpha_0 \gg 1 $ and   
$   |\tilde{E} \tilde{V} |\ll \alpha_{0}  $ 
from (\ref{75}) the eigenvalues \cite{Barbier1, Dietel3}  
 \begin{equation} 
\tilde{\epsilon}_\pm^0  = \! \!\pm  ( s \hat{\alpha}^2_0 
 \sqrt{\tilde{k}^2_x + 
|\hat{\Gamma}|^2 \tilde{k}_y^2}- \tilde{\mu})      
   \,  .         \label{80} \\
\end{equation}
Here $ \hat{\Gamma}=\sin[\alpha_{0}]e^{i \alpha_0}/\alpha_{0} $, 
$ \hat{\alpha}_0=\alpha_0/\tilde{V}$. 
We define dimensionless quantities $ \tilde{x} \equiv x d/2 
\hbar v_F $ for quantities $ x $ having the dimension of energy and  
$ \tilde{k} \equiv  k d/2 $ when $ k $ has as an inverse length dimension. 
The  
Bloch momentum in $ x $-direction is restricted to 
$ -\pi/2 \le \tilde{k}_x \le \pi/2 $. The parameter  
$ s=1 $ denotes the conduction band and $s=-1$ the valence band. 
We show in the left panel 
in Fig.~1 the approximation to the lowest lying energy band 
$ \tilde{\epsilon}_+^0 $ 
(\ref{80}) (solid curves) and its exact counterpart 
(dotted curves) at $ k_x=0 $ and $ \mu=0$, $s=1$ 
for various SL potentials $ \tilde{V} $. 
We obtain a good agreement between both curves except at the outer boundary 
of the folded region 
where $ \tilde{k}_y/\tilde{V} \approx 1 $. Here we find  
$ |\tilde{E}\tilde{V}|/\alpha_0 \sim 1 $ close to the VHSs, implying a  
breakdown of the expansion.  The solution 
$ u^{0}_{\pm}(0) $ is given in the regime $   
|\tilde{E} \tilde{V}|  \ll \alpha_0, \alpha_{0}\gg 1  $ by 
\begin{equation} 
u^{0}_{\pm}(0) \approx \left(
\begin{array}{c} 
\frac{ \cos(\alpha_0) \sin(\alpha_0)}{\alpha_0} \tilde{k}_y  +i 
\tilde{k}_x   \\
i \frac{1}{\hat{\alpha}_0^{2}}\, \tilde{E}_\pm  + 
i \frac{\sin^{2}(\alpha_0)}{\alpha^2_0} \tilde{V}    \tilde{k}_y       
\end{array} \right)\,.   \label{100} 
\end{equation}
We shall denote the vector components by 
$ u^{0}_{\pm}(0)=(A+i\tilde{k}_x,iB)^T $. 
From (\ref{80}) we obtain an oscillatory behavior of the lowest energy band 
as a function of $ k_y $. New Dirac points emerge 
at $ {\bf k}=0 $ for $ \tilde{V} \in  \mathbb{N}  \pi $. We compare 
in Fig.~1 Eq.~(\ref{80}) with a numerical solution of (\ref{70}).  
The new Dirac points are shifted along the y-axis in $ {\bf k}$-space 
for increasing $ \tilde{V} $.
Now we focus on the higher energy saddle points building 
singularities in the density of states. The figure shows that even in 
this energy regime the approximation (\ref{80}) is justified. 
Saddle points are quite interesting 
in forming a high-temperature BCS state when the chemical potential 
is close to the VHS. 
By using (\ref{80}), we obtain for the density  
of states $ \nu(\epsilon)$ per spin and valley close to a VHS at energies 
$ E^n_+ =\epsilon^0_+(0,k^n_y)+\mu $, originating from a saddle point 
with momentum $ k_y=k^n_y $ and $k_x=0 $ for $ \alpha_{0}\gg 1  $
\begin{align} 
& \nu(\epsilon)  \approx  \frac{\tilde{\nu}_0}{\hbar v_F d} 
\ln\left(\frac{16 \tilde{W}^2_{\rm VHS}}{|(\tilde{\epsilon})^2-
(\tilde{E}_+^n)^2|} \right)\,,  \label{110}  \\
& 
\tilde{\nu}_0 = \frac{\sqrt{2}}{\pi^2} 
\frac{|\tilde{E}_+^n| \tilde{V}^4}{(\tilde{k}^n_y)^2 \alpha^2_0}
\frac{1}{\sqrt{\frac{1}{2}+ \cos^2( \alpha_0)}} \,, 
 \nonumber  
\end{align} 
where $ \tilde{W}_{\rm VHS}= {\rm min}[\hat{\alpha}^2_0 \pi/2, 2 |\tilde{\mu}|] $ 
is the width of the VHS.   
We obtain from (\ref{80}) the relation $ \tan(\alpha_0) \approx 
(\tilde{k}^n_y)^2/\alpha_0 $
for the momentum $ \tilde{k}^n_y $ of the $ n $-th  saddle point 
in the energy spectrum where $ n=1 $ corresponds to  
the outermost saddle point. The 
solution of $ \tan(\alpha_0) \approx 
(\tilde{k}^n_y)^2/\alpha_0 $ can be  
approximated for the outer saddle points  
by $ \tilde{k}^n_y  \approx \pm \sqrt{\tilde{V}^2-(\pi/2+ n \pi)^2} $  
for $ n \in \{1,\ldots,[\tilde{V}/\pi]-1 \} $. Here $ [x] $ 
is the largest integer value smaller than $ x $. 
The saddle point closest to 
the central Dirac point has then still to be determined numerically by 
 $ \tan(\alpha_0) \approx (\tilde{k}^n_y)^2/\alpha_0 $.  

Due to the oscillatory behavior of the energy band 
we obtain that even for small chemical potentials, electrons with energies 
close to the chemical potential can scatter 
with a large momentum transfer. This  
is relevant when determining 
the energy cut-off within BCS theory, which we denoted $ \omega_{\rm DB}^* $. 
By using (\ref{40})--(\ref{70}) 
with (\ref{100}) we obtain that the lowest band wavefunctions are  
localized around the canonical momenta $ k_x \approx  \pm \alpha_02/d $ and
$ k_y $. This then leads to the energy cut-off   
for acoustic $ {\bf \Gamma} $-phonon scattering 
$\omega_{\rm DB}^* \approx  {\rm min}[V/\hbar v_F k_{\rm DB},1] \omega_{\rm DB}$.

As it was mentioned in the introduction, the energy 
cut-off for graphene without an SL due to acoustic electron-phonon scattering  
is in general much smaller being, 
$ \omega_{\rm DB}^* \approx  (\mu/\hbar v_F k_{\rm DB})\omega_{\rm DB} $.

\section{One-particle spectrum}

By using (\ref{30}) we are now able to calculate the $ \Delta $ dependent 
correction terms to $ {\bf \Lambda} $.  
With the abbreviation  $ \hat{\bf \Delta}(x)= (-\sigma_3 \otimes \sigma_0 \otimes \sigma_3) \cdot (\sigma_1\otimes \Delta) $ we obtain  
\begin{align} 
& {\bf \Lambda}^1(x)=i {\bf \Lambda}^0(x)\int^x_0 dx'
({\bf \Lambda}^0)^{-1}(x') \hat{{\bf \Delta}}(x')  {\bf \Lambda}^0 (x')  \,,  \label{120} \\
 & {\bf \Lambda}^2(x)=- {\bf \Lambda}^0(x)\int^x_0 dx'
({\bf \Lambda}^0)^{-1}(x') \hat{{\bf \Delta}}(x') {\bf \Lambda}^0 (x') 
\label{122} \\ 
&  \qquad \qquad \times \int^{x'}_0 dx''({\bf \Lambda}^0)^{-1}(x'') \hat{{\bf \Delta}}(x'')
{\bf \Lambda}^0 (x'')  \,. 
  \nonumber 
\end{align}
Here we use $ {\bf \Lambda}^0=\sum_{p \in \{+,-\}}  
(\sigma_0+p \sigma_3)\otimes \sigma_0 \otimes \Lambda^0_p/2 $.   
In the following, we calculate perturbationally 
the eigenvalues of the transfermatrix $ {\bf \Lambda}(d) $ where  
$ {\bf \Lambda}^1 $ and $ {\bf \Lambda}^2 $ are seen as perturbations to  
$ {\bf \Lambda}^0 $.  

We point out that standard Rayleigh-Schr\"odinger 
perturbation theory is not applicable  here since 
the transfer matrices $ {\bf \Lambda}^0 $ 
or $ {\bf \Lambda} $, respectively, are neither unitary nor hermitian. 
This is due to the fact that the matrix on the right hand side in 
Eq.~(\ref{30}) is not hermitean. But this matrix is hermitian with respect 
to the quadratic form $ \langle {\bf u}|{\bf v}\rangle_Q \equiv 
 \langle \sigma_3\otimes \sigma_0 \otimes \sigma_3  {\bf u}|
{\bf v}\rangle $. Thus {\it it does } lead  to the unitarity of 
$ {\bf \Lambda} $ and  ${\bf \Lambda}^0 $ with respect to this form. 
Note that this  
quadratic form is not positive definite. In the Bloch regime where 
the eigenvalues are a pure phase factor,  
different eigenvalues are orthogonal with respect to the 
$ Q$-form. One can now show that standard Rayleigh-Schr\"odinger 
perturbation  
can be used after all by substituting the quadratic form 
$ \langle {\bf u}|{\bf v}\rangle_Q $ for all expressions 
where normally the cartesian scalar product 
$ \langle {\bf u}|{\bf v}\rangle $ is used. 
This includes also the normalization of the basis 
functions (\ref{100}). 

In order to calculate the matrix elements of the operators (\ref{120}), 
(\ref{122})  
we have 
taken into account the degeneracy of the eigensystem of  
$ {\bf \Lambda}^0 $. 
In zero's order perturbation we obtain 
degenerate eigenstates. 
With the abbreviation $ e_i $ ($i=1,\ldots,4$) for the cartesian basis 
in four-dimensional space, we obtain for the eigenvectors  
of $ {\bf \Lambda}^0 $,    
$  u_+^0 \otimes e_1 $, $  u_+^0 \otimes e_2 $ with eigenvalues of either  
$ e^{+ik^+_xd} $ and $e^{-ik^+_xd} $ in the particle sector and  
$  u_-^0 \otimes e_3 $, $  u_-^0 \otimes e_4 $ with eigenvalues   
$ e^{+ik^-_xd} $ and $e^{-ik^-_xd} $ in the hole sector. 
Here $ k_x^\pm $ is given by $ \tilde{k}_x^\pm = \sqrt{ \tilde{E}_\pm^2/\hat{\alpha}^4_0
- |\hat{\Gamma}|^2 \tilde{k}_y^2} $. Note that $ {\bf \Lambda}^1 $ and 
$ {\bf \Lambda}^2 $ can in first approximation only connect states 
which are in the lowest band, i.e. $ \tilde{k}^\pm_x  \le \pi/2 $.    
By denoting $ M_{\pm V}=k_y \sigma_3 \mp V\sigma_2 / \hbar v_F $ we obtain 
for $ \alpha_0 \gg 1 $ 
\begin{align} 
& {\bf \Lambda}^1(d) \approx i \frac{d}{4} \bigg[{\bf \Lambda}^0(d) 
 \hat{\bf \Delta}_{V}\left(\frac{d}{4}\right) \!\!  + \! 
\hat{\bf \Delta}_{-V}\left(\frac{3d}{4}\right) {\bf \Lambda}^0(d)\bigg], 
\label{125} 
\\ 
& {\bf \Lambda}^2(d) \approx -\frac{d^2}{16}\bigg(\frac{1}{2}
\bigg\{{\bf \Lambda}^0(d) \left[\hat{\bf \Delta}_{V}\left(\frac{d}{4}\right)
\right]^2 
\!\!\! + \! 
\left[\hat{\bf \Delta}_{-V}\left(\frac{3d}{4}\right)\right]^2  \nonumber \\
&\quad \quad \quad  \times {\bf \Lambda}^0(d)\bigg\}   
+\hat{\bf \Delta}_{-V} \! \left(\frac{3d}{4}\right) 
\! {\bf \Lambda}^0(d) \hat{\bf \Delta}_{V} \! \left(\frac{d}{4}\right)\bigg)
 \label{130} 
\end{align}    
 where 
$ \hat{\bf \Delta}_{\pm V}= \hat{\bf \Delta}- M_{\pm V} \hat{\bf \Delta} 
M_{\pm V} $.   
Next we calculate the matrix elements of ${\bf  \Lambda}^{1}(d) $,  
${\bf  \Lambda}^{2}(d) $ with respect to the basis $ u^{0}_{\pm} \otimes e_i $. 
Here we can restrict ourselves to leading order in $ \epsilon^0 $ and 
$ k_x^\pm $ justified for chemical potentials close to a VHS. 
We obtain $ u^{0} \equiv u_-^{0} \approx u_+^{0} $ with $ \Lambda_{ij} =
\langle  u^{0} \otimes e_i |{\bf \Lambda}^1(d)| u^{0} \otimes e_j \rangle_Q $  for $i = 1, \ldots,  4 $ 
\begin{align}
& {\Lambda}_{31}^1(d)={\Lambda}_{13}^1(d)={\Lambda}_{24}^1(d)={\Lambda}_{42}^1(d) \nonumber \\
& 
\quad  \quad  \approx 
- \frac{4 \tilde{V} [d^c_1 \tilde{V} 
 (A^2+B^2+ \tilde{k}^+_x \tilde{k}^-_x)+2 d^s_1
\tilde{k}_y A B ] }{\tilde{V}^2-\tilde{k}_y^2} ,    \nonumber \\ 
 &  {\Lambda}_{32}^1(d)=-{\Lambda}_{23}^1(d)={\Lambda}_{14}^1(d)=-{\Lambda}_{41}^1(d)  \,, \label{140} \\
& 
\quad  \quad  \approx 
i \frac{4 \tilde{k}_y [d^s_3 \tilde{V} 
 (A^2+B^2+ \tilde{k}^+_x \tilde{k}^-_x)+2 d^c_3
 \tilde{k}_y A B ] }{\tilde{V}^2-\tilde{k}_y^2} \,. \nonumber          \end{align} 
In contrast to (\ref{140}), the matrix elements $ {\Lambda}_{ij}^2(d) \not=0 $
are much more complicated, being also a function on the condensates 
$ d_2 $ with prefactors similar to (\ref{140}).  
We even include in (\ref{140}) a 
subleading $ \tilde{k}^2_x $-term, which becomes relevant for 
the $ d_1, d_3 $ dependence of the spectrum when the 
$ k_y $ momentum lies not close to the saddle point.  

To zero's order in $ \Delta $,  
we find two different $ \tilde{\epsilon}^0 $-regimes
within Rayleigh-Schr\"ordinger perturbation theory.   
For small $ |\tilde{\epsilon}^0| $ where $ k_x^{+} \approx k_x^{-} $ 
we find approximately 
a fourfold degenerate ground state with momentum 
$ k_x^{+} $ in the $ {\bf K} $ and $-{\bf K} $ valleys 
in the electronic sector, and 
$ k_x^{-} $ in the $ {\bf K} $ and $-{\bf K} $ valleys 
in the hole sector. The same holds 
for the $- k_x^{\pm} $ momenta. 
This degeneracy is lifted by using $ {\bf \Lambda}^1(d) $
within first order perturbation theory. The 
energy spectrum is then dominated by the first order energy with respect 
to $ {\bf \Lambda}^1(d) $.  

For larger $ |\tilde{\epsilon}^0| $ where $ k_x^{+} \not=k_x^{-} $, we find a
 two-fold degeneracy corresponding to the $ k_x^{+} $-state in the 
$ {\bf K} $, $ -{\bf K} $ electron valleys and a further 
degenerate ground state with $ k_x^{-} $ in the 
$ {\bf K} $, $ -{\bf K} $ hole valleys. The same holds 
for the $- k_x^{\pm} $ momenta.
The electron and hole valleys are 
not degenerate with each other 
in this case. The degeneracy for small $|\tilde{\epsilon}^0|$ where 
$ |\tilde{\mu} \tilde{\epsilon}^0| \ll 1 $ 
is lifted by first order 
perturbation theory with respect to  $ {\bf \Lambda}^2(d) $ in this case. 
On the other hand the 
first order energies with respect to $ {\bf \Lambda}^2(d) $ can 
be neglected in comparison to the second order energies with respect to 
$ {\bf \Lambda}^1(d) $.  

To simplify our condensate search further, we will first consider only the  
$ d_1 $ dependence of the energy spectrum setting $ d_i=0 $ for $ i \not=1 $.   
By taking into account the consideration following (\ref{70}) we obtain 
$ {\rm Tr}[\Lambda_\pm(d)]+ D_\pm =2 \cos(k_x d) $ with 
 \begin{align} 
&  D_\pm \approx \pm {\rm min}
  \left[
  \frac{|T_1|}{4 B}, 
  \frac{|T_1|^2 }{
  64 B^2 |\tilde{\mu} \tilde{\epsilon}_\pm^0| \hat{\alpha}_0^4} 
  \right] 
           {\rm sgn}[\tilde{\mu} \tilde{\epsilon}^0_\pm]
\Theta(\pi/2-\tilde{k}^\mp_x)  \, , 
 \label{170} 
  \\
   &  T_1= - \frac{4 \tilde{V}}{\alpha_0^2}[\tilde{d}_1^c 
\tilde{V} (A^2+B^2+ \tilde{k}^2_x) + 2 
  \tilde{d}_1^s\tilde{k}_y  AB] \,.
 \label{180}        
 \end{align} 
With the help of (\ref{80}) we obtain for the branch of the energy spectrum 
being mainly influenced by BCS pairing for 
$  |\tilde{\epsilon}^0| \ll |\tilde{\mu}| $   
\begin{align} 
& \tilde{\epsilon}_\pm \approx \pm 
\left(s \sqrt{(\tilde{\epsilon}_\pm^0 \pm \tilde{\mu})^2+
D_{\pm}/ \hat{\alpha}^4_0} -\tilde{\mu} \right) \nonumber \\   
& \quad  \approx \pm {\rm sgn}[\tilde{\epsilon}^0_+] 
\sqrt{(\tilde{\epsilon}^0_+)^2 
+ D^2}\,,  \label{190}
\end{align} 
with $ D=D_c(d^c_1,d^s_1) $ where 
\begin{equation} 
D_c(d^c_1,d^s_1)= \frac{1}{\hat{\alpha}_0^4} \frac{|T_1|}{8 B \tilde{\mu}}
\Theta(\pi/2-\tilde{k}^-_x)
\,.  
\label{200}
\end{equation}  
 Note that in (\ref{190}) with (\ref{80}), 
the band parameter $ s $ has to be chosen such that 
$ |\tilde{\epsilon}^0_+| \ll |\tilde{\mu}| $, 
i.e. $ s={\rm sgn}[\tilde{\mu}] $.  
The energy bands  in  (\ref{190}) are  doubly degenerated.
This degeneracy is lifted when going 
beyond the lowest approximation used here. 

The energy spectrum (\ref{190}) with (\ref{200}) has now a similar form 
as the energy spectrum of metals within the standard BCS theory.
This point can be elaborated further by taking into 
account that (\ref{10}) with (\ref{20}) 
where only $ d_1^c \not=0 $ but $ d_1^s =0 $ and $ d_i^c,d_i^s =0 $ for 
$ i \not=1 $, can be diagonalized by using standard Bogoljubov theory. 
This is based on the fact that 
$ \Delta $ is comuting with $ H_+ $. This leads to the energy spectrum 
(\ref{190}) with (\ref{200}) where now $ D_c= \tilde{d}_1^c$. 
This means that we should find  $ D_c(1,0) \approx 1 $ in expression 
(\ref{200}) in order to have a good approximation in hand. 

We show in Fig.~1 $ D_c(1,0) $ for various 
SL potentials $ \tilde{V} $ and chemical potentials $ \tilde{\mu} $ 
as a function of the rescaled momentum 
$ \tilde{k}_y/\tilde{V}$ (left inset) and $ \epsilon^+_0=0 $. 
The curve segments which are absent in the figure are  
where  $ 0 \le \tilde{k}_x^+\le \pi/2 $ 
is not fulfilled. Right panel in Fig.~1 shows 
$ D_c(1,0) $ and $ D_c(0,1) $ calculated at $ \tilde{k}_y $-momenta 
and chemical potentials $ \tilde{\mu} $ of the saddle point for the 
VHS singularities 
$n=1,\ldots, [\tilde{V}/\pi] $. 
We obtain from the figure or (\ref{100}), respectively, 
that for large $ \tilde{V} $ and small $ n $ (outer VHSs), 
$ D_c(1,0) $ is growing to infinity which can be avoided 
by taking into account higher order corrections in 
$ \tilde{E} \tilde{V} /\tilde{\alpha}_0$ in (\ref{100}) 
(c.f. caption of Fig.~1).
From the right panel in Fig.~1, we obtain that the largest 
 $ D_c(0,1) $ value is reached for the outermost VHS with $n=1 $ where 
$ \tilde{V}\approx 4 $ with value $  D_c(0,1) \approx 0.3 $. 
A further exceptional SL potential for $ n=1 $ is given by 
 $ \tilde{V}= 6.66 $ where 
$ D_c(0,1) $ is vanishing. We show in the right inset in Fig.~1 
the energy spectrum $ \tilde{\epsilon}_+ $ as a function of $ d^s_1 $ for 
$ d^c_1=0  $ using these both exceptional SL potentials and 
further the SL potential $ \tilde{V}=4 \pi $ ($n=1 $)  
to gain a better insight what is happening with  the spectrum in the 
outer VHSs for large $ \tilde{V} $. We compare our results in the figure 
with a numerically determined energy spectrum for the same values 
using a numerically evaluated 
transfermatrix method similar to (\ref{30})--(\ref{70}).  

Summarizing we obtain from Fig.~1 
that the agreement of our approximations with exact and numerical 
results are good for small $ \tilde{V} \gtrsim  1 $
but also for $ \tilde{V} \gg 1 $ for the inner valleys. 
The approximation becomes  less good for the outermost valleys. 
The reason lies in the expansion parameters 
$ 1/\alpha_0 $ and $ \tilde{E}  V/\alpha_0 $ which we used 
in our approximation in
order to derive (\ref{190}), (\ref{200}).    

\begin{figure*}
\begin{center}
\includegraphics[clip,height=7.5cm,width=14.0cm]{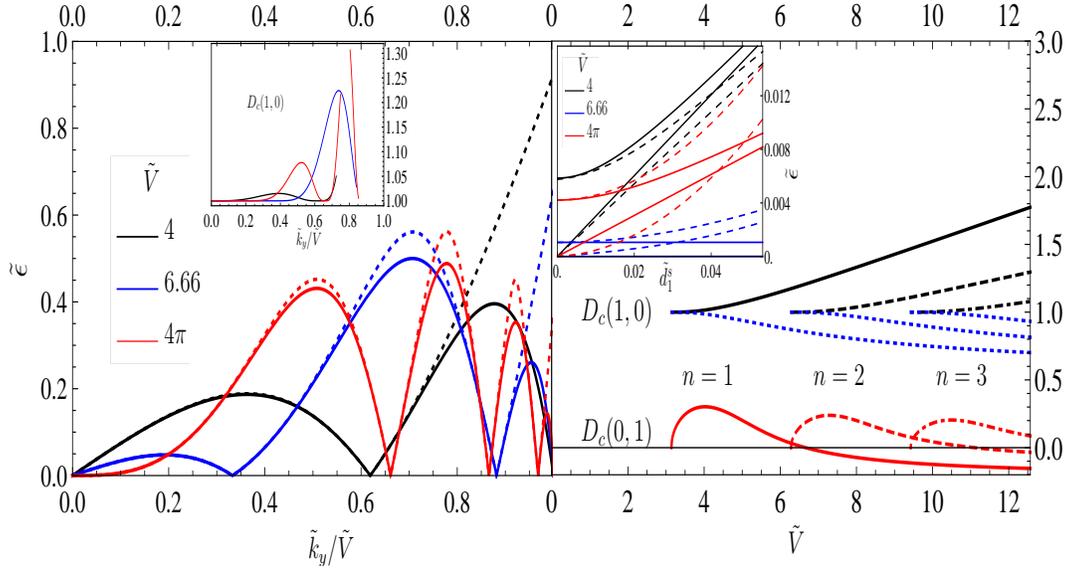}
\vspace{-2em}
\end{center}
\caption{Left panel: Energy spectrum $ \epsilon^0_+ $ 
(\ref{80}) at $ \tilde{k}_x=0 $ and $ \tilde{\mu}=0 $, $s=1 $ 
for various SL potentials $ \tilde{V} $. Dotted curves show 
the corresponding exact spectrum obtained 
by evaluating the transfer matrix eigenvalue 
equation (\ref{70}) numerically. Inset shows $ D_c(1,0) $ (\ref{200}) 
as a function of $ \tilde{k}_y/\tilde{V} $ 
for $ \epsilon^0=0 $ and SL potentials $ \tilde{V}=4, 6.66, 4 \pi $.  
The curves are calculated by using the outer valley VHS chemical 
potentials $ \tilde{\mu} = 0.185, 0.5$ in the case $ \tilde{V}=4, 6.66$, and 
by using the average chemical potential of the three existent VHSs 
 $ \tilde{\mu} = 0.415$ being of similar absolute energy value for 
$ \tilde{V}=4\pi $. 
Right panel: $ D_c(1,0) $ and $ D_c(0,1) $ for 
$ \epsilon^0 = 0$, $ k_y=k_y^n $ 
 and chemical potentials $ \tilde{\mu} $  at the VHSs 
$ n=1, \ldots, 3 $ where $ n=1 $ corresponds to the outermost VHS.
The dotted curves show $ D_c(1,0) $ by going one order 
higher taking into account 
(\ref{100}) up to order $ (\tilde{E} \tilde{V}/ \alpha_0)^2 $. 
$ D_c(0,1) $ is not changed within this approximation.    
Inset shows the energy spectrum 
$ \tilde{\epsilon}_+ $ (\ref{190}) 
for $ d_1^c=0 $ as a function of $ \tilde{d}_1^s $ for $ \epsilon^0=0 $ and 
also one further value $ \epsilon^0 \not = 0$. The specific value can be 
read off from the intersection of the spectral curve with the $ y$-axis. 
The corresponding dashed curves are calculated by a numerical  
diagonalization of (\ref{10}) using a transfer matrix method similar 
to (\ref{30})--(\ref{70}). }  
\end{figure*} 

Until now, we have only discussed the $ d_1 $-dependence of the energy 
spectrum. From Eq.~(\ref{140}), we obtain that close to a VHS 
for pure condensates, i.e. where  
$ d_i\not=0 $ for only one $ i $ and the rest of the condensates is 
zero,  only the $ d_3 $ beside the $ d_1 $ 
condensate has a nonzero contribution in the gap function $ D $. The 
$ d_2 $-dependence in the gap function comes in via $ \Lambda^2_{ij} $, 
leading to mixing terms  of the pure condensate contributions 
to the gap function.   That the 
$ d_4 $-condensate does not contribute to the gap function is caused by 
the fact that $ \hat{{\bf \Delta}}_{\pm V} $ does not depend on 
$ d_4 $.

For the $ d_3 $ dependence of the energy gap function $ D $, i.e. by setting  
$ d_i=0 $ for $ i \not=3 $,  we obtain the expression (\ref{200}) with the 
substitutions $ d_1^c \rightarrow d_3^s $, $ d_1^s \rightarrow d_3^c $,  
and after a multiplication of a reduction factor 
$ \tilde{k}_y/\tilde{V}$.  
The reduction factor has its origin in the prefactor 
differences between $ \Lambda^1_{31} $ and 
$ \Lambda^1_{32}$  (\ref{140}). 
In general, we obtain for the energy spectrum (\ref{190})  
in the relevant large energy regime
$ |\tilde{\epsilon}^0| \gg {\rm max}[|\tilde{d}_i|] $ for superconductivity  
$ D^2=D^2_{c,i} $ where  
\begin{align} 
& D^2_{c,i} =D^2_c(d_1^c,d_1^s)+ \frac{\tilde{k}^2_y}{\tilde{V}^2}
D^2_c(d_3^s,d_3^c)   \label{205} \\
&
\qquad - 2 \frac{\tilde{k}_y}{\tilde{V}} D_c(d_1^c,d_1^s)
D_c(d_3^s,d_3^c) \frac{2 {\rm Im}[EV_i]}{1+|EV_i|^2} \,, 
\nonumber 
\end{align}
and $ i=1,2$. 
Here we denoted by $ (1,EV_i)^T $ for $ i=1,2 $ 
as the eigenvectors of the matrix $ \Lambda^2_{ij} $ for $ i,j \in \{1,2\} $ 
and  
$ EV_i $ is a function of the condensates $ d_1, d_2$, and $d_3 $.
In the less relevant regime 
$ |\tilde{\epsilon}^0| \ll {\rm max}[|\tilde{d}_i|] $, 
the gap function $ D $ looks similar where $ 2 {\rm Im}[EV_i]/(1+|EV_i|^2)=
\pm 1 $. 
We now obtain from (\ref{205}) that the degeneracy of the  
energy spectrum seen for the pure $ d_1 $ condensates in  (\ref{190}) 
with (\ref{200}) is lifted.       

\section{BCS-instability} 
We are now able to  
calculate from the one-particle spectrum (\ref{190}) 
the $ \Delta $-dependent part of the 
grand canonical potential $ \tilde{\Omega} $.  
The condensates $ d_i $ are  then determined by minimizing 
$ \tilde{\Omega} $ with respect to the 
pair functions $ d_i^s,d_i^c $. We restrict our search of 
the minimum thereby by comparing the minimum of the free energies in the 
various basic directions where $ d_i \not=0 $ for one $ i $ but zero for 
the others. This restriction is justified by taking into account the 
the smallness of the condensate mixing term in Eq.~(\ref{205}) 
and further that the energy 
regime $ |\tilde{\epsilon}^0| \gg {\rm max}[|d_i|] $ in the spectrum gives the 
dominant contribution to the free energy integral in the weak coupling regime 
(see the discussions below). For the 
mixing last term in (\ref{205}) we mention  
that $ 2 {\rm Im}[EV_i]/(1+|EV_i|^2) \le 1$ is strongly dependent 
on the momenta and condensate values  $ d_1, d_2, d_3 $. For a justification 
of its smallness 
one can show that $ 2 {\rm Im}[EV_{i}]/(1+|EV_{i}|^2) $   
is zero for $ d_2=0 $ and becomes 
much smaller than one at least for one $ EV_i $ in
 the regime where $ d^c_1\sim d_3^s \gg d^s_1, d^c_3 $. 

When considering only the large energy regime  
together with the neglecting of the mixing term, 
our restricted minimum search in the free energy is then even exact. 
Due to the additional small prefactor 
$ \tilde{k}_y/\tilde{V} $ of the condensate contributions of 
$ d_3$ in comparison to $ d_1 $ in the energy gap $ D $ the 
condensate $ d_1\not=0 $ leads to a smaller free energy than the 
other condensates. This results in the free energy          
\begin{align}
& \tilde{\Omega} \frac{d^2}{S} \! = \! -
\frac{32 \tilde{T}}{(2 \pi)^2}\frac{1}{\hat{\alpha}^4_0} 
 \! \! \! \! \! \!  \!\! \!  \! \int\limits_{0}^{\sqrt{\tilde{V}^2-(\pi/2)^2}} 
\! \! \! \! \! \! \! \! \! \! d\tilde{k_y} 
 \int\limits_{-\tilde{\omega}_{DB}^*}^{\tilde{\omega}^*_{DB}}  \! \! \! \! d \tilde{\epsilon}^0_+ 
\bigg\{ \Theta(\pi/2-\tilde{k}_x^+) 
\frac{|\tilde{E}_+|}{\tilde{k}_x^+}  \nonumber \\
& \times \log
\left[2+2 \cosh\left(\frac{\tilde{\epsilon}_+}{\tilde{T}}\right) \right] 
\bigg\} 
+\frac{16}{\tilde{g}}  \, [(\tilde{d}_1^c)^2+(\tilde{d}_1^s)^2]\,    \label{210} 
\end{align}   
where $ \tilde{g}=2 g  /d \hbar v_F $.  
The condensate values  $ d_1^c $, $ d_1^s $ are  then determined by minimizing 
$ \tilde{\Omega} $. We show in Fig.~2 the resulting  
$ d_1^c $, $ d_1^s $ values as a function of the dimensionless 
temperature $ \tilde{T} $ for various SL potentials $ \tilde{V}$.
The dimensionless effective Debye frequency is given by 
$ \tilde{\omega}_{\rm DB}^* \approx 0.017 \tilde{V} 
{\rm min}[d/a \tilde{V},1] $.  
In Kelvin we obtain, assuming a maximal longitudinal acoustic
phonon frequency in graphene of $ \omega_{\rm DB} \approx 
1960 K $,  
$ \omega_{\rm DB}^* \approx 1960 {\rm K}  
\cdot {\rm min}[\tilde{V} a/d ,1] $. 
From Fig.~1 we obtain that the highest 
critical temperatures $ T_c $ are gained for large 
$ \tilde{V} $. For $ \tilde{V}=4 \pi $, ($\tilde{V}=6.66$),  
(($\tilde{V}=4 $)) we obtain 
$ T_c=\{1136 {\rm K}, 315 {\rm K}, 58 {\rm K}\} $ ($ 
\{587 {\rm K}, 104 {\rm K}, 9 {\rm K}\}$) (($\{76 {\rm K}, 2.5 {\rm K}, 0.154 
{\rm K}\}$)) at $ d/a= \{8, 16, 32\} $. 
We find further $ \tilde{T}_c/\tilde{\omega}_{\rm DB}^*=\{0.58,0.205,0.075\} $,  
($ \{0.36,0.13,0.02\}$), (($ \{0.078,0.005 ,0.0006\}$)) and    
$ \tilde{d}^c_1/\tilde{\omega}_{\rm DB}^* \approx \{1.06,0.35,0.1\} $,
($\{0.62,0.21,0.017\}$), (($\{0.13,0.0196,0.0011\}$)) at $T=0 $.
It is well known that due to decoherence effects of the electronic 
wavefunction for $ T_c/\omega_{\rm DB}^*\gtrsim 1  $ and the neglection 
of retardation in the Fr\"ohlich Hamiltonian for 
$ d_1/\omega_{\rm DB}^*\gtrsim 1  $, 
the BCS results cannot be trusted any
longer in this regime. The regime 
is commonly called the  intermediate to strong-coupling regime.   
$ T_c $ as well as  $ d_1 $ are then truncated at $ \omega_{\rm DB}^* $ 
\cite{Alexandrov1}. A better description in this  regime takes into
account  higher order fluctuation effects as well as 
the frequency dependence of the 
effective electron-electron interaction being described by Eliashberg 
theory in the intermediate coupling regime and polaron superconductivity 
for strong couplings \cite{Alexandrov1}. The results in both regimes 
for metals as well as for pristine graphene within  Eliashberg theory
\cite{Einenkel1} 
suggest that a realistic cut-off for $ T_c $ should be in the vicinity 
of $ \omega_{\rm DB}/3 $, leading to $ T_c $ values up to $ 650 {\rm K} $. 

The analysis in this  paper is based on the effective mass approximation 
(\ref{5}) for the graphene Hamiltonian. This approximation is justified in the 
case of the linearity of the graphene spectrum. The linearity 
is fulfilled in first approximation 
for momenta $ |{\bf k}| \lesssim k_{\rm BZ}/2 $ around the 
$ {\bf K} $, $ -{\bf K} $ points. 
The relevant $ k_y $-momentum  
of a saddle point of a VHS calculated by (\ref{5}) should then lie in this 
momentum regime. This regime is roughly fulfilled for the parameters of the 
SL potentials shown in  
Fig.~2. We point out that
the whole analysis in the last two sections 
is mainly based on the folding behavior  of the energy band. 
This behavior is a much more stable property with respect to perturbations
of the graphene lattice  than for example the creation of new Dirac 
points. This justifies further the use of the effective mass approximation 
for VHSs with saddle points at large effective momenta. 
     
In the low coupling regime ($T_c \ll W $), 
we obtain from (\ref{210}) by using (\ref{110}) 
\begin{equation} 
\log^2\left(\frac{\tilde{W}^2_{\rm VHS}}{\tilde{T}_c |\tilde{\mu}|/2} 
\right) 
-\log^2\left(\frac{\tilde{W}^2_{\rm VHS}}{ \tilde{W}|\tilde{\mu}|/2} 
\right) \approx  \frac{16}{\tilde{\nu}_0 \tilde{g}} \frac{1}{\|
D_c\|^2} \,, 
\label{220} 
\end{equation}      
where $ \tilde{W} $ is the effective band width 
$ \tilde{W}= {\rm min}[\tilde{\omega}_{\rm DB}^*,\tilde{W}_{\rm VHS}] $
 and $ \left\| D_c \right\| \equiv \sqrt{D_c(1,0)^2+
D_c(0,1)^2}$ calculated with the saddle point momentum $ k_y=k_y^n $.   
The condensates $ \tilde{d}_1^c $, $ \tilde{d}_1^s $ at $ T=0 $ 
in the low-coupling regime are given by 
\begin{equation} 
\tilde{d}^c_1= D_c(1,0)  \| \tilde{d}_1 \| 
/\| D_c \|, 
\tilde{d}^s_1= D_c(0,1) \| \tilde{d}_1 \| /
\| D_c \| \,. \label{225} 
\end{equation} 
Here $ \| \tilde{d}_1 \| $ is given by (\ref{220}) with the substitution 
$ 2 \tilde{T}_c \rightarrow 
\| D_c \| \| \tilde{d}_1 \| $. 
In the strong coupling regime $ \tilde{W} $ 
replaces $ \tilde{\omega}_{\rm DB}^* $ as a cut-off for $ \tilde{T}_c $ and 
$ \| \tilde{d}_1\| $ for $ T=0 $.

From (\ref{220}), (\ref{225}),  we obtain then that in leading order 
$ d^s_1/d^c_1 \approx D_c(0,1)/D_c(1,0)$ calculated for $ k_y=k_y^n $ at 
$ T=0 $. This is qualitatively in accordance 
to Fig.~2 by using the results for $ D_c(0,1) $, $ D_c(1,0) $ 
in Fig.~1. By this we mean that  
$ d^s_1/d^c_1 $ is much larger for $ \tilde{V}=4 $ in comparison to 
$ \tilde{V}= 4 \pi $, $6.66 $.  
Nevertheless we obtain quantitatively discrepancies 
which are attributed to contributions in the gap equation 
(\ref{210}) which are not taken into account by the 
VHS contribution (\ref{220}). 
For a justification we mention  
that $ D_c(0,1)/D_c(1,0) $ is oscillatory as a function of $ k_y $.  
For example for 
$ \tilde{V}=4 $ we obtain that $ D_c(0,1)/D_c(1,0) \approx 0.3 $ at 
$ k_y=k_y^1 $. This is almost the maximum value of $ D_c(0,1)/D_c(1,0) $ 
as a function of $ k_y $, showing even negative values 
$ D_c(0,1)/D_c(1,0) \approx -1 $ for larger $ k_y $.             
   
\begin{figure*}
\begin{center}
\includegraphics[clip,height=5.5cm,width=17.0cm]{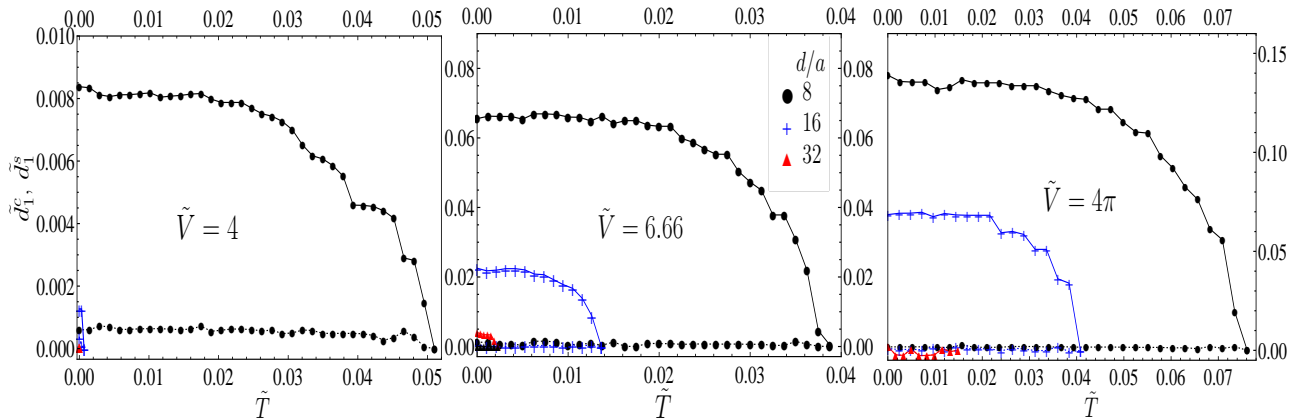}
\vspace{-2em}
\end{center}
\caption{We show the condensate quantities $ \tilde{d}^c_1 $ 
(upper solid curves) and $ \tilde{d}^s_1 $ (lower dotted curves)
for various SL potentials $ \tilde{V} $ as a function of temperature
by minimizing the free energy (\ref{210}).  
The chemical potentials are chosen to lie at the outer 
valley VHSs for $ \tilde{V}=4, 6.66 $ and 
at the average of the VHS energies for $ \tilde{V}=4 \pi $ (see caption 
to Fig.~1)}  
\end{figure*} 

Finally we compare our results with the phonon mediated 
superconductivity in pristine graphene without an SL. 
We shall calculate in the following  
$ T_c $ for acoustic phonon pairing and in a rough approximation 
also for optical phonon pairing in order to demonstrate the 
proportion of the critical temperatures for both pairing mechanisms. 
We restrict ourselves  hereby to the $ d^c_1 $ pairing mechanism which leads to  
\begin{equation} 
\log\left(\frac{\hbar \omega^p_{\rm DB}}{T_c}\right) \approx 2 \frac{1}{g_p \nu_p(\mu)}
\,. \label{230} 
\end{equation}  
The cut-off frequency 
$ \omega^p_{\rm DB} $ is given by 
$ \omega^p_{\rm DB} \approx \omega_{\rm DB} $ for optical 
phonon pairing and $ \omega^p_{\rm DB} \approx (|\mu|/\hbar v_F k_{\rm DB}) 
\omega_{\rm DB}$ for 
acoustic phonon pairing. For the former we use that the acoustic Debye frequency 
and the optical phonon frequencies are of similar value in graphene 
\cite{Suzuura1}.   
The density of states $ \nu_p(\mu) $ per spin and valley 
for pristine graphene is given by 
$\nu_p(\mu) = |\mu|/2\pi (\hbar v_F)^2 $. The constant $ g_p $ in (\ref{230}) 
is the effective Fr\"ohlich 
interaction constant being $g_p= g \approx 6 \cdot 10^{-19} m^2{ \rm eV}$ 
for acoustic phonon pairing and 
$ g_p \approx 7.02 \cdot 10^{-20} m^2 {\rm eV}  $ for pairing with 
optical phonons 
\cite{Calandra2}. The factor two on the right-hand side of 
Eq.~(\ref{230}) is attributed to the chiral nature of the graphene lattice
with two atoms in the fundamental cell where   
for large chemical potentials only electrons in one of the bands 
$ \pi^* $ or $ \pi $ with energies close to the chemical potential 
can pair. The maximal absolute electron density which can be reached 
by electrostatic doping till now leading to the highest 
$ T_c $ values is given by 
$ n_e \approx 4 \cdot 10^{14} cm^{-2} $ \cite{Efetov1}. 
By using (\ref{230}) for this density 
we obtain $ T_c \approx  4.1 \cdot 10^{-3} \omega_{\rm DB} 
\approx 8 {\rm K}$ for acoustic phonon coupling 
(here $ |\mu|/\hbar v_F k_{\rm DB} \approx 1/5 $) and
$ T_c \approx 2.7 \cdot 10^{-15} 
\omega_{\rm DB} \approx 5.4 \cdot 10^{-12} {\rm K} $ for optical phonons. 
These transition temperatures are much smaller than most of the transition 
temperatures in graphene superimposed by an SL with parameters used in Fig.~2.

Until now, we have restricted our minimum search of the free energy 
to  condensates of the form (\ref{20}) 
showing the full symmetry of the SL together with the time inversion symmetry
and spin singlet form. 
In general the condensate matrix $ \Delta $ has no restrictions 
from the beginning.   
The BdG Hamiltonian (\ref{10}) shows 
an independent chiral symmetry in the electronic and hole sector. 
We are justifying in App.~A the utilized condensates (\ref{20})  
by showing that the condensate  
$ d_1^c $ modulo its chiral symmetric counterparts, 
i.e. $ \Delta \rightarrow (U_1^+\otimes \sigma_0) \Delta
 (U_2 \otimes \sigma_0)  $ where $ U_1, U_2 $ 
are arbitrary constant unitary 
$ 2 \times 2 $  matrices and $\Delta=\Delta_1 $, 
have the largest condensate 
values together with the minimal free energy 
and dominate the BCS pairing process. We use hereby 
as was implicitly also used above that 
the Fr\"ohlich coupling constant $ g $ for acoustic phonons
is not depending on the pairing deduced from $ H_{\rm ep}$.  
This is  not fulfilled for other  coupling mechanisms as for example 
the coupling with optical phonons.
A benefit of the analysis  used in App.~A is that it can be simply adapted 
to other coupling mechanisms.  

It is well known that in two dimensions the phase fluctuations of a continuous 
order parameter are so strong that 
a finite order parameter value calculated in mean-field 
vanishes in higher order approximations 
(Hohenberg-Mermin-Wagner-Theorem). Nevertheless 
a finite expectation value for the amplitude of the order parameter is 
still possible. At lower temperatures where the order parameter amplitude 
is non-zero a Kosterlitz-Thouless transition emerges which is connected 
to an unbinding of vortex-antivortex excitations when crossing the temperature 
from below \cite{Kleinert1, Minnhagen1}. The free vortices 
prohibit then in the so called pseudo-gap phase 
that a true superconductivity behavior is existent. 
At lower temperatures where the vortices are bound, 
we can find in two-dimensional systems superconductivity.
In other words, the mean-field BCS theory which we formulated 
in this paper, approximatively can only describe  
the transition temperature where the pairing amplitude 
is unequal to zero being then an upper bound for the true superconducting 
phase transition temperature. This temperature difference 
where the order parameter amplitude becomes unequal to zero  
and where the vortex unbinding happens 
is not large at least in the regime where  
$ \mu \gg \omega_{\rm DB} $ was shown quantitatively 
in the case of two-dimensional metals in 
Refs.~\onlinecite{Loktev1, Loktev2} by using 
Eliashberg theory. Due to this, we also expect in the case of 
the graphene system that the two temperatures are quite close to each other.

\section{Summary} 
We have examined possible BCS instabilities mediated by acoustic 
$ {\bf \Gamma} $ phonons in electrical superlattice systems. Here we 
restrict ourselves to SL potentials $ \tilde{V} \gtrsim  1 $, $ d/a \gg 1 $  
where $ \tilde{V} a/d  $ is not too small such that the 
acoustic phonon coupling is in fact the dominating phonon coupling process.  
In the regime $ \tilde{V} \gtrsim  1 $, the energy bands are folded 
where new Dirac points linked by 
low-lying energy VHSs emerge. We considered in this paper mainly pairing 
for chemical potentials close to  VHSs where the highest  $ T_c $ 
temperatures are attained. For SL systems such chemical potentials 
should be reached by electrostatic doping. 
We showed under the assumption of a pairing that 
fulfills time inversion symmetry together with the symmetry 
of the SL and graphene lattice that electronic onsite s-wave pairing 
of an electron around the $ {\bf K} $ point 
with another electron around the $ -{\bf K} $ point is most relevant. The 
relevant order parameter is almost constant in space. 
We obtain large transition temperatures $ T_c $ especially where VHSs 
lie close to each other. 
We have compared the calculated $ T_c $ values of the SL system 
with phonon mediated transition temperatures
of electrostatic doped pristine graphene.   
Finally, we argued that the encountered order parameter 
(up to chiral symmetry) is also  
the leading electronic pairing mechanism when taken 
into account no symmetry restrictions on the condensate matrix.

We have used in this paper the simplest theory for superconductivity 
appropriate for pairing in the low coupling limit for electrons around 
the $ {\bf K}$, $ -{\bf K} $ points. Our examples in Fig.~2 produce 
superconductivity at rather high temperatures, and at the highest 
$ T_c$ values  the system parameters lies at the 
validity boundary of the model. In this case, the calculated $ T_c $-values
are only a rough approximation for the experimental 
transition temperatures where more exact calculations 
would be useful by using for example the full tight binding Hamiltonian 
together with Eliashberg theory for the SL superimposed  graphene system.

\appendix
\section{The dominance of $ d^c_1 $ condensates and its chiral equivalences among general condensates}
  
In the main text we considered only highly symmetric condensates
as possible electron pairings which fulfill the full 
mirror symmetry of the SL and additional time inversion 
symmetry and spin singlet pairing. This led to the condensates (\ref{20}) 
as the only contributions to the matrix $ \Delta $.  
As was mentioned in the main text,  we have in general no 
restriction for acoustic phonon 
coupling on the condensate matrix 
$ \Delta $. In the following we shall 
use again the approximation that the matrix is step-like in space meaning 
that it is constant 
for constant $ V(x) $. 
In weak-coupling BCS physics, the regime $ |\epsilon^0_\pm| \gg 
{\rm max}[|\Delta_{ij}|] $ of the spectrum is most relevant for  
superconducting pairing. 
Let us recall from the main text in Sect.~III that in the case of 
the highly symmetric condensates, the dominance of the 
$ d^c_1 $-condensate contributions over the $ d^s_1 $- and 
$ d_3 $-contributions came mainly 
from the fact that in the gapfunction $ D_c $ 
(\ref{200}) the prefactor  $ \sim A^2+B^2 $ 
for $ d_1^c $ is much larger than the prefactor $ \sim A B $ 
for $ d_1^s $. Furthermore we found 
the dominance of the $ d_1 $ condensates over the 
$ d_3 $ condensates due to an additional prefactor 
$ \tilde{k}_y/\tilde{V} $ in the $ d_3 $ condensate term (\ref{205}).   

These prefactors were calculated by using (\ref{125}) 
with (\ref{100}). Within a similar argument we obtain 
that the dominant contributions for general $ \Delta_{ij} $ are given by 
$ \Delta^d= \sum^3_{i=0} 
do_i \sigma_i \otimes \sigma_0 $. 
The condensates $ do_i $ are in general complex and constant over the whole SL. 
Other condensates of the matrix form $ \sum^3_{i=0} do_i \sigma_i \otimes \sigma_1 $ and  $ \sum^3_{i=0}  do_i \sigma_i \otimes \sigma_3 $ 
lead to energy gap contributions 
being a factor   $A B/(A^2+B^2) $ smaller where condensates 
of the form $ \sum^3_{i=0} do_i  \sigma_i \otimes \sigma_2 $ are a factor 
$ \tilde{k}_y/\tilde{V} $ smaller. 

By using the chiral invariance of (\ref{10}) 
in the electron and hole sector independently 
we can restrict ourselves by using the singular value decomposition of the matrix 
$ \Delta^d $ to matrices $ \Delta^d =do^r_1 (\sigma_3+\sigma_0)/2\otimes 
\sigma_0+ 
do^r_2 (\sigma_3-\sigma_0)/2 \otimes \sigma_0$. 
Here $ do^r_1 $ and $ do^r_2 $ are real 
condensates being constant over the SL. 
The dominant mass gap contributions $ D= D^{d}_{c,i} $ are then given by 
\begin{equation} 
  D^{d}_{c,i} = D_c(1,0)
\sqrt{\frac{(do^r_1)^2 + |EV_2|^2 (do^r_2)^2}{1+|EV_2|^2}}\,.   \label{a10}
\end{equation} 
Here  $ (1,EV_1)^T $, $ (1,EV_2)^T $ are the two orthogonal eigenvectors 
of the matrix  $ \Lambda^2_{ij} =
\langle  u^{0} \otimes e_i |{\bf \Lambda}^2(d)| u^{0} \otimes 
e_j \rangle_Q $ for $ i,j=1,2 $ 
in the electronic sector where now also 
contributions from smaller subleading condensate contributions can have 
a strong influence via $ EV_i $ on the free energy. 
For deriving (\ref{a10}) we took into account the discussion 
following (\ref{140}). Note that the spectrum (\ref{a10}) 
with (\ref{190}) for $ D= D^{d}_{c,1} $ and $ D= D^{d}_{c,2} $, respectively, 
is now in general no longer degenerate as 
in (\ref{170})-(\ref{200}) but has two nondegenerate bands 
with two different gap values. 
We now minimize the dominant part of the free energy 
first with respect to $ EV_1 , EV_2$.  The $ EV_i $-dependence 
comes then in only via the first term in (\ref{210}) where now we have to 
substitude $ \log[2+2\cosh(\tilde{\epsilon}_+/\tilde{T})] $ by 
$ \sum_i \log[2+2\cosh(\tilde{\epsilon}^i_+/\tilde{T})]/2 $. Here  
$ \tilde{\epsilon}^i_+ $ is defined via (\ref{190}) using (\ref{a10}) 
with the substitution $ D^{d}_{c,i} $ for $ D $ . 
By using the concavity  of 
$ \log[2+2\cosh(\sqrt{(\epsilon^0_+)^2+x}/ \tilde{T})] $ as a function of 
$ x $, and further 
that $ (D^{d}_{c,1})^2+ (D^{d}_{c,2})^2 $ does not depend 
on $ EV_1 $ and $ EV_2 $ we obtain that the minimal free energy is attained for 
$ |EV_1| = |EV_2|=1 $. The dominant 
contribution to the free energy $ \tilde{\Omega}^d $ 
is then given by (\ref{210}) with the substitutions above 
where we further have to substitude $ (\tilde{d}^c_1)^2 +
(\tilde{d}^s_1)^2 $ by $ ((\tilde{do}^r_1)^2+ (\tilde{do}^r_2)^2)/2 $. 
This free energy shows a $ O(2) $ invariance. By choosing   
$ do^r_1= -do^r_1 $ we obtain exactly the $ d^c_1 $-contribution 
to the condensate matrix (\ref{20}).

\end{document}